# Eightfold Fermionic Excitation in a Charge Density Wave Compound

Xi Zhang,[1,2,*] Qiangqiang Gu,[1,2,*] Haigen Sun,[3,*] Tianchuang Luo,[1,2,*] Yanzhao Liu,[1,2] Yueyuan Chen,[1,2] Zhibin Shao,[3] Zongyuan Zhang,[3] Shaojian Li,[3] Yuanwei Sun,[1,4] Yuehui Li,[1,4] Xiaokang Li,[3,5] Shangjie Xue,[1,6] Jun Ge,[1,2] Ying Xing,[7] R. Comin,[6] Zengwei Zhu,[3,5] Peng Gao,[1,2,4,8] Binghai Yan,[9] Ji Feng,[1,2,†] Minghu Pan,[3,‡] and Jian Wang[1,2,8,10,§]

[1]*International Center for Quantum Materials, School of Physics, Peking University, 100871 Beijing, China*
[2]*Collaborative Innovation Center of Quantum Matter, 100871 Beijing, China*
[3]*School of Physics, Huazhong University of Science and Technology, 430074 Wuhan, China*
[4]*Electron Microscopy Laboratory, School of Physics, Peking University, 100871 Beijing, China*
[5]*Wuhan National High Magnetic Field Center, Huazhong University of Science and Technology, 430074 Wuhan, China*
[6]*Department of Physics, Massachusetts Institute of Technology, Cambridge, 02139 Massachusetts, USA*
[7]*Department of Materials Science and Engineering, School of New Energy and Materials, China University of Petroleum, 102249 Beijing, China*
[8]*Beijing Academy of Quantum Information Sciences, 100193 Beijing, China*
[9]*Department of Condensed Matter Physics, Weizmann Institute of Science, 7610001 Rehovot, Israel*
[10]*CAS Center for Excellence in Topological Quantum Computation, University of Chinese Academy of Sciences, 100190 Beijing, China*

[*]These authors contributed equally to this work.
[§]Corresponding author.
jianwangphysics@pku.edu.cn (J.W.)
[‡]Corresponding author.
minghupan@hust.edu.cn (M.P.)
[†]Corresponding author.
jfeng11@pku.edu.cn (J.F.)

**Abstract**

Unconventional quasiparticle excitations in condensed matter systems have become one of the most important research frontiers. Beyond two- and fourfold degenerate Weyl and Dirac fermions, three-, six- and eightfold symmetry protected degeneracies have been predicted however remain challenging to realize in solid state materials. Here, charge density wave compound TaTe$_4$ is proposed to hold eightfold fermionic excitation and Dirac point in energy bands. High quality TaTe$_4$ single crystals are prepared, where the charge density wave is revealed by directly imaging the atomic structure and a pseudogap of about 45 meV on the surface. Shubnikov de-Haas oscillations of TaTe$_4$ are consistent with band structure calculation. Scanning tunneling microscopy reveals atomic step edge states on the surface of TaTe$_4$. This work uncovers that charge density wave is able to induce new topological phases and sheds new light on the novel excitations in condensed matter materials.

## I. INTRODUCTION

Condensed matter systems, as translational symmetry broken universe, host passionately pursued quasiparticle excitations analogous to high-energy elementary particles. A successful example is the prediction and observation of Dirac [1-3] and Weyl [4-6] fermions in solids protected by band topology. Weyl semimetals (WSMs) hold twofold degenerate Weyl points with nontrivial Chern number in energy bands. Fourfold degenerate Dirac points (DPs) with linear dispersion protected by time-reversal and inversion symmetry exist in Dirac semimetals (DSMs) and are stabilized by uniaxial rotation symmetries [7] or nonsymmorphic space group (SG) symmetries [1]. Recently, relation between Dirac or Weyl fermions and collective phenomena such as superconductivity (SC) is addressed [8-11], which is regarded as a route to realize topological SC and condensed matter analogue of a third type of elementary particle, Majorana fermions [12].

However, the marvelous story is to be continued. First, classification of condensed matter fermionic excitations by Dirac, Weyl and Majorana is incomplete. The 230 SG symmetries in three dimensional (3D) lattices impose constraints weaker than Poincaré symmetry required in quantum field theory and therefore permit free fermionic excitations with no high-energy analogue. Complete enumeration predicts three-, six- and eightfold degenerate fermions in condensed matter systems in addition to Dirac and Weyl fermions [13,14]. Threefold degeneracy has already been observed in MoP [15] and WC [16] while other types are still awaiting confirmation in real materials.

Second, apart from SC, another collective phenomenon of particular interest, the charge density wave (CDW), is much less studied in the context of topological materials. CDW is the ground state of a wide range of bulk compounds showing quasi-one dimensional (quasi-1D) or quasi-two dimensional (quasi-2D) lattice structures such as $NbSe_3$ [17], $(TaSe_4)_2I$ [18,19] and some transition metal dichalcogenides [20] arising from electron-phonon interaction. When cooling below Peierls transition temperature $T_p$, periodic modulation of charge density as well as structural distortion occur in these materials and an energy gap is opened at Fermi level. Similar to SC, CDW is also characterized by a complex order parameter $\Delta e^{i\varphi}$, where the phase $\varphi$ is of fundamental importance, whose fluctuation forms gapless current-carrying phase mode. However, phase mode of CDW can be pinned by impurities, lattice defects and, in the commensurate case, lattice itself. As a result, they can only give rise to nonlinear conductivity above a threshold electric field $E_T$ [21].

Interplay between CDW and topology has been theoretically considered in the context of 2D Dirac semimetals [22], topological insulators (TIs) [23], topological WSMs [24,25] and DSMs [26]. In week TI, CDW is predicted to gap the surface states, however, create helical edge states on CDW domain walls [23]. In WSMs and DSMs, interaction effect may induce CDW which gaps topologically protected degeneracy point and the phase of CDW couples to electromagnetic field in an analogous way to axion [25,26]. However, up to now, experimental studies of CDW in topological semimetals are insufficient to test these hypotheses. Alternatively, attempts of identifying topological phases in typical CDW compounds are still lacking.

In this article, we report topological semimetal phase in a CDW compound $TaTe_4$. $TaTe_4$ is structurally regarded as one of the simplest inorganic materials hosting CDW [27]. Our calculation further points out that $TaTe_4$ is a topological semimetal both with and without the presence of CDW. In its undistorted crystal structure P4/mcc (SG124) without CDW, $TaTe_4$ holds DPs along high symmetry lines $\Gamma - Z$ and $M - A$. A commensurate CDW phase in $TaTe_4$ is calculated to be the zero-temperature ground state and hold double Dirac point (DDP) with eightfold degeneracy at A as well as a Dirac point at Z in CDW Brillouin zone. The calculated band structure of CDW phase is consistent with the observed Shubnikov de-Haas (SdH) oscillations. The nontrivial topology of $TaTe_4$ is also suggested by the scanning tunneling spectroscopy (STS) detection of states residing on atomic step edges on the surface of $TaTe_4$. The coexistence of CDW and fermionic excitations has been rarely reported before in other materials. Our results reveal the CDW-induced topological phases, which may lead to the possibility of dynamically tuning the topological properties of crystalline materials, and provide new insight into the interplay between CDW and fermionic excitations.

## II. EXPERIMENTAL AND CALCULATION METHODS

### A. Band structure calculations

Density functional theory (DFT) calculation is performed using Vienna Ab initio Simulation Package (VASP) [28] within the generalized gradient approximation, parameterized by Perdew, Burke, and Ernzerhof [29]. The energy cutoff for a plane-wave basis is set to 400 eV and the *k*-point mesh is taken as $11 \times 11 \times 11$ for non-CDW phase and $5 \times 5 \times 5$ for CDW phase. The CDW structure of TaTe$_4$ is obtained by fully relaxing a $2a \times 2b \times 3c$ supercell of the non-CDW phase of TaTe$_4$ after adding a random distortion. The fully relaxation procedure was performed with a conjugated gradient algorithm, until Hellmann–Feynman force on each atom is less than $10^{-3}$ eV/Å. The spin-orbit coupling effect is self-consistently included.

### B. Single-crystal growth

High quality TaTe$_4$ single crystals are prepared by chemical vapor transport (CVT) technique in a multi-zone single crystal furnace [30]. Ta and Te powder are mixed at atomic ratio 1:4.3 and sealed with 5-10 mg/cm$^3$ iodine in an evacuated quartz tube. The mixture was first heated to 1000 °C and kept for 12 hours. After the required period of heating, sample is slowly cooled down to room temperature. The quartz tube is then placed co-axially in a multi-zone furnace. The reaction zone is kept at 540 °C and growth zone 440 °C for 7 days. After slowly cooling down to room temperature, bulk samples with metallic luster are obtained. Samples grown by this method are quite stable in atmosphere.

### C. Sample characterization and transport measurements

Chemical composition of prepared crystals is confirmed by energy dispersive X-ray spectroscopy (EDS). The HAADF STEM image is performed on an aberration-corrected transmission electron microscope FEI Titan Cubed Themis G2 300. The single crystal property of the sample is confirmed by Bruker D8 Advance powder X-ray diffractometer. $\rho - T$ measurement and magnetotransport measurement under magnetic field up to 15 T is performed in a Physical Property Measurement System (PPMS-16 T) by Quantum Design. Magnetoresistance up to 56 T is measured at Wuhan National High Magnetic Field Center with a pulsed magnetic field. Standard 4-probe/6-probe configuration is used throughout transport measurements.

### D. Scanning tunneling microscopy (STM) measurements

For STM study, the crystals chosen approximately of the size 2 mm×1 mm×0.5 mm. Samples are first cleaved *in situ* at room temperature in an ultrahigh vacuum chamber with pressure lower than $1 \times 10^{-10}$ torr. After cleavage, the crystal exhibits a mirror-like surface. An electro-chemically etched tungsten wire is used as the STM tip. All measurements by STM were conducted at liquid helium temperature 4.2 K.

## III. RESULTS AND DISCUSSION

### A. Crystal structure and schematic electronic structure of TaTe$_4$

TaTe$_4$ crystalizes in tetragonal P4/mcc space group in the absence of CDW. Quais-1D chains along *c* axis are formed by every Ta surrounded by 8 Te atoms [Fig. 1a] (also see Fig. S1 in Supplementary Material (SM) [31]). At room temperature, TaTe$_4$ exhibits commensurate CDW phase, which enlarges its unit cell to be $2a \times 2b \times 3c$ [Fig. 1a] (also see SM [31] Fig. S2), where *a*, *b* and c are lattice vectors in non-CDW unit cell. With CDW distortion, Ta atoms form Ta$_3$ clusters in the chain while Te atoms are slightly rotated around the Ta chains, which develop the equivalent 1D chains into three distinct types. The simulated CDW phase crystal structure is consistent with previous and our experimental observations [38,39]. The space group under CDW phase in P4/ncc where C4 rotation and inversion symmetry are preserved. We prepare TaTe$_4$ single crystals using chemical vapor transport technique (details are presented in Methods section). The atomically resolved high-angle annular dark-field (HAADF) scanning transmission electron microscopy (STEM) image manifests the high quality of our sample [Fig. 2a]. Schematic structure clearly shows lattice distortions. After performing fast Fourier transform (FFT), four spots around the center are detected to represent enlarged unit cell in CDW phase (SM [31] Fig. S3a). XRD is also performed for *ac* surface. Its result further confirms a uniform CDW phase in our sample (SM [31] Fig. S3b).

The band structure corresponding to non-CDW structure of TaTe$_4$ is shown in Fig. 1b. Two DPs are detected and marked by the red dashed rectangle in Fig. 1b near the Fermi level. The Dirac point DP1 lies in the Γ − Z line and 293 meV below the Fermi level, while the Dirac point DP2 lies 288 meV above the Fermi level in the M-A line. The Fermi energies of DPs (the energies of DPs relative to the Fermi level) in non-CDW TaTe$_4$ are comparable to that in Cd$_3$As$_2$ [40,41]. Fig. 1c shows the band structure of TaTe$_4$ after considering CDW induced lattice distortion. The two DPs in non-CDW band no longer exist. Instead, an eightfold degeneracy (or a DDP) at A point protected by symmetries of SG130 arises [13,14] at 644 meV below the Fermi level [Fig. 1c and d]. Also, a new Dirac point located 457 meV below the Fermi level appears at Z point [Fig. 1c]. As a result, CDW dramatically modifies band structure and Fermi surface (FS) of TaTe$_4$ and TaTe$_4$ holds eightfold fermionic excitation in CDW phase. Our calculation results indicate that CDW may strongly affect the topological property of a material, which has not been seriously considered before.

Interestingly, a surface CDW phase distinct from bulk CDW is observed by STM. At 4.2 K, topographic image of cleaved *ac* surface measured by STM shows clear periodic modulation [bright strips in Fig. 2c]. FFT of image shows CDW peak at around $\pm 2\pi(\frac{1}{4a}, 0, \frac{1}{6c})$ (SM [31] Fig. S3d), indicating an enlarged $4a \times 6c$ surface unit cell which is larger than bulk unit cell by CDW lattice distortion ($2a \times 2b \times 3c$) and breaks C$_4$ rotation symmetry on the surface. Intriguingly, for the region where CDW modulation is present, a pseudogap of approximately 45 meV is observed around Fermi level, as is shown in Fig. 2d. This is interpreted as the CDW gap [42] since such gap is reduced on the surface region where CDW modulation is less ordered (SM [31] Fig. S4).

### B. Magnetoresistance and magnetic oscillations in TaTe$_4$

The resistivity-temperature ($\rho - T$) relation of TaTe$_4$ is shown in Fig. 2b. Sample resistivity exhibits metallic behavior under zero magnetic field and saturates to around 3.5 μΩ·cm below 10 K, with a rather large residual resistivity ratio (RRR = $\frac{\rho(300 \text{ K})}{\rho(2 \text{ K})}$) 111.

Magnetotransport measurement setup is schematically plotted in the lower inset of Fig. 2b. Current is always along $c$ axis and magnetic field is rotated relative to the sample in $ab$ plane. When magnetic field is applied along the $a$ axis ($\theta = 0°$), longitudinal magnetoresistance defined by MR = $\frac{\rho(B) - \rho(B=0)}{\rho(B=0)} \times 100\%$ saturates to around 3800% at 15 T [Fig. 3a] while pulsed magnetic field measurement up to 56 T reveals more complicated structure of MR at higher fields [Fig. S5a-c]. When magnetic field is applied along $\theta = 45°$, MR does not fully saturate up to 15 T [Fig. 3c]. At both magnetic field directions, oscillations of MR are clearly noticeable. By subtracting smooth background, $1/B$ periodic SdH oscillation components are obtained [43,44], as are shown in Fig. 3b and Fig. 3d. Oscillation frequencies can be obtained by performing FFT analysis of the data. The results for FFT of oscillations at lowest temperature are shown in the inset of Fig. 3b and Fig. 3d respectively. For $\theta = 0°$, two oscillation frequencies, labeled by $\beta$ and $\beta'$ are observed, along with the second harmonic of $\beta$. We identify $F_\beta = 330$ T and $F_{\beta'} = 509$ T. They originate from equivalent pockets, as will be shown later. By Onsager relation [45] and assuming a spherical FS cross section, we obtain Fermi wave vector $k_F = 0.10$ Å$^{-1}$ for $\beta$ pocket. For $\theta = 45°$, aside from frequency $F_\beta = 386$ T close to that observed for $\theta = 0°$ and its harmonics, another smaller frequency $F_\alpha = 34$ T is observed. $F_\alpha$ corresponds to $k_F = 0.031$ Å$^{-1}$. As a result, SdH oscillation pattern seems to be the superposition of two kinds of oscillations showing different frequencies. Additionally, a very high frequency (1383 T) is also detected. This kind of very high frequency oscillations is further revealed by ultrahigh magnetic field experiments (SM [31] Fig. S5d-f) which may originate from magnetic breakdown effect [46].

Often used for SdH oscillation analysis is a simplified version of the standard Lifshitz-Kosevich (LK) formula [45,47]:

$$\Delta MR \propto R_L R_D \cos\left[2\pi(\frac{F}{B} + \phi)\right] \tag{1}$$

where $R_L = \frac{\chi}{\sinh(\chi)}$, $\chi = \frac{2\pi^2 k_B m_c T}{e\hbar B}$ describes the damping of oscillation amplitude with temperature and $R_D = \exp\left(\frac{\pi m_c}{eB\tau_Q}\right)$ is the Dingle factor. $m_c$ is the cyclotron mass and $\tau_Q$ the quantum lifetime. Analysis of SdH oscillation patterns provides rich information on the electronic structure of TaTe$_4$ in CDW phase, which tests the validity of our calculations. By fitting oscillation amplitude as a function of temperature (SM [31] Fig. S7a), we obtain cyclotron mass $m_c = 0.33 m_e$ for $\beta$ pocket at $\theta = 0°$, where $m_e$ is the electron mass. Fermi velocity is also obtained as $v_F = 3.5 \times 10^5$ m/s. Further evidences [Fig. 4b] show that pocket $\beta$ is of three-dimensional (3D) nature with rather low anisotropy. Thus, we can estimate the contribution to carrier concentration of a single $\beta$ pocket by $n_\beta^{pocket} =$ $\frac{2}{(2\pi)^3}\frac{4}{3}\pi k_F^3 = 3.4 \times 10^{19}$ cm$^{-3}$, which is comparable with that obtained by Hall measurement at 2 K (SM [31] Fig. S8) $n^{\text{Hall}} = 1.34 \times 10^{20}$ cm$^{-3}$. For $\alpha$ pocket at $\theta = 45°$, $m_c = 0.15\, m_e$ (SM [31] Fig. S7b) and $v_F = 2.32 \times 10^5$ m/s. Since $\alpha$ pocket is highly anisotropic [Fig. 4c], we cannot reasonably estimate its contribution to carrier concentration.

Angular-dependent magnetoresistance [48] or magnetic oscillation is a powerful tool to study FS anisotropy. Offset plot of FFT of oscillations at different $\theta$ (SM [31] Fig. S9) is shown in Fig. 4a, where we can extract peak frequencies corresponding to $\alpha$ and $\beta$ pockets. $F_\alpha$ is most prominent around $\theta = 45°$ and $\theta = 135°$ and can hardly be identified around $\theta = 0°$, 90° and 180°, while $F_\beta$ can be detected at all angles. Both $F_\alpha$ and $F_\beta$ show repeated pattern every 90°, again consistent with the $C_4$ lattice symmetry. The experimentally detected frequencies can be identified by the extremal orbits in calculated CDW phase FS, with calculated frequencies presented as violet points in Fig. 4b and c. Four bands cross the Fermi level of TaTe$_4$ in CDW phase, among which band 1 and 2 give rise to open FS while band 3 and 4 exhibit closed Fermi pocket at the Fermi level, as is shown in Fig. 4d-g. Band 2 does not stably form a closed orbit and thus is hard to be detected by magnetic oscillation. In comparison, although FS of band 1 is extended, it holds cylindrical sections along Z − A line, which may explain the quasi-2D behavior of $F_\alpha$ (red lines in Fig. 4c). With these considerations, we assign $F_\alpha$ to band 1 and $F_\beta$ to band 4 to obtain best fit to our experimental data. The calculated effective mass $m^*$ in this scenario is $0.21\, m_e$ for $F_\beta$ at 0° and $0.10\, m_e$ for $F_\alpha$ at 45°, which is also in agreement with the cyclotron mass obtained by fitting the temperature dependence of oscillation amplitude ($m_c = 0.33\, m_e (\beta)$ and $0.15\, m_e (\alpha)$ respectively). Therefore, the calculated band structure can explain the transport properties of CDW phase TaTe$_4$ well, which validates our calculation. In addition, the non-CDW phase band structure cannot fit experimentally observed frequencies (SM [31] Fig. S10), further confirming the CDW modulation on energy bands.

### C. States residing on atomic step edges on the surface of TaTe$_4$

When a topological semimetal is reduced to two dimensions, 1D topological edge states could be detected. To further confirm the topological property of TaTe$_4$, we carry out STM measurement to study the edge of TaTe$_4$ terraces. The cleaved *ac* surfaces with steps of 2a height are shown in Fig. 5a and b. Such step edges can exist with weak [Fig. 5a] or clear [Fig. 5b] CDW modulation. Interestingly, edge states are detected under both circumstances at different biases. Fig. 5a shows an edge existing in a region without clear CDW modulation, which is mostly parallel to Ta chain direction (*c* axis). We acquire tunneling differential conductance spectra, proportional to local density of states (LDOS), along a line across a step edge. Tunneling conductance has finite value for all energies, further demonstrating the metallic properties revealed by transport measurement. On approaching the step edge, tunneling conductance shows a drastic increase at around -75 meV within the range of 2 nm [Fig. 5c], similar to that observed in WTe$_2$ [49]. Such increase in LDOS is detected at several different edges at the same energy and thus cannot be well explained by a trivial edge state. In regions with clear CDW, edge state appears at around 12 meV and coexists with CDW gap [Fig. 5d]. The presence of edge state makes the CDW gap seems to be narrower. These results show that TaTe$_4$ can hold edge states with or without clear CDW in 2D limit, which not only suggests the topological nature of TaTe$_4$ but also points out that it is altered by the presence of CDW.

### D. Discussions

CDW has been studied as a low dimensional phenomenon in various materials. Typically, it features a gap opened at Fermi level, which is identified as a transition behavior in $\rho - T$ and insulating behavior below $T_{\rm p}$, as in the case of $(TaSe_4)_2I$ [18], $K_{0.3}MoO_3$ [50] and $TaS_3$ [51]. As for quasi-1D material $NbSe_3$, it remains metallic at low temperatures since one type of $NbSe_3$ chain remains metallic [17]. Above scenarios cannot be applied to $TaTe_4$, since CDW in $TaTe_4$ is commensurate and all chains exhibit trimerization with fixed relative phase [39]. According to our STS results, CDW does not fully gap the Fermi surface of $TaTe_4$, which indicates that under the CDW phase, 1D and 3D features exist simultaneously and only part of the electrons are involved in the CDW phase, as is also concluded by previous angle-resolved photoemission spectroscopy (ARPES) measurement [52]. The remaining DOS at Fermi level contributes to the sample conductance at low temperatures.

As a typical quasi-1D CDW compound, $TaTe_4$ is gifted to hold exotic fermionic excitation as revealed by our DFT calculations. More interestingly, the fermionic excitations are altered by CDW, which is not expected for typical quasi-1D compounds with CDW transition whose FS is fully gapped by CDW. The calculated eightfold degenerate point in $TaTe_4$ is closer to Fermi level than the ARPES observed triply degenerate point in MoP [15], so that the related novel properties can be more easily detected. Direct observation of the DDP and DP in $TaTe_4$ by ARPES technique is highly desired. The Dirac line nodes associated with the DDP in $TaTe_4$ may give rise to pairs of drumhead surface states [13]. Also, Dirac line nodes serve as a parent phase of all known topological semimetals. External symmetry-breaking perturbations, such as a magnetic field or uniaxial strain, may lead to DPs, Weyl points or twofold degenerate line nodes [13]. Moreover, the existence of DP and DDP in $TaTe_4$ at different momenta may enable study of interplay between different kinds of fermions.

What is more interesting about TaTe$_4$ is that it also serves as a platform to study the interplay between CDW and fermionic excitations. Experimentally, CDW has rarely been studied in the context of topological semimetals. In proposed Weyl semimetal candidate Y$_2$Ir$_2$O$_7$, CDW was found to gap out the possible Weyl nodes [53]. The origin of CDW in Y$_2$Ir$_2$O$_7$ is quite unusual and may be related to chiral symmetry breaking and axion dynamics. Theoretically, interaction between Weyl fermions at different momenta can induce CDW, which gaps Weyl or Dirac points and the phase mode is predicted to present itself as axion in condensed matter systems [25,26]. Considering the robustness of CDW in TaTe$_4$, axion dynamics could be observed in TaTe$_4$ single crystal above the critical electrical field. Although the small resistance and the commensurate nature of CDW hinders the investigation of collective mode's contribution to the conductance of bulk TaTe$_4$, the nanoscale samples with higher resistance might offer the chance to detect axion dynamics and optical methods could also be helpful [21]. Another promising way is to investigate Nb doped TaTe$_4$, which is found to tune the CDW phase from commensurate to incommensurate [27,54]. Further investigations on the interplay between CDW and topology in TaTe$_4$ are highly desired.

## IV. CONCLUSIONS

In summary, our results show that CDW phase TaTe$_4$ is a topological semimetal with DDPs as well as DPs. The calculated band structure of TaTe$_4$ is consistent with the observed SdH oscillations. STS detection of states residing on atomic step edges on the surface of TaTe$_4$ further indicates the nontrivial topology of TaTe$_4$. Therefore, our work suggests that TaTe$_4$ is a promising platform to study physical phenomena related to CDW and free fermionic excitations in condensed matter systems.

## ACKNOWLEDGEMENTS

This work was financially supported by the National Key R&D Program of China (Grant Nos. 2018YFA0305600 and 2017YFA0303302), the National Natural Science Foundation of China (Grant Nos. 11888101, 11774008, 11574095, 11725415, 51672007), the Strategic Priority Research Program of Chinese Academy of Sciences (Grant No. XDB28000000), and Beijing Natural Science Foundation (Z180010). We gratefully acknowledge Gatan for the technique help. The authors acknowledge Electron Microscopy Laboratory in Peking University for the use of Cs corrected electron microscope.

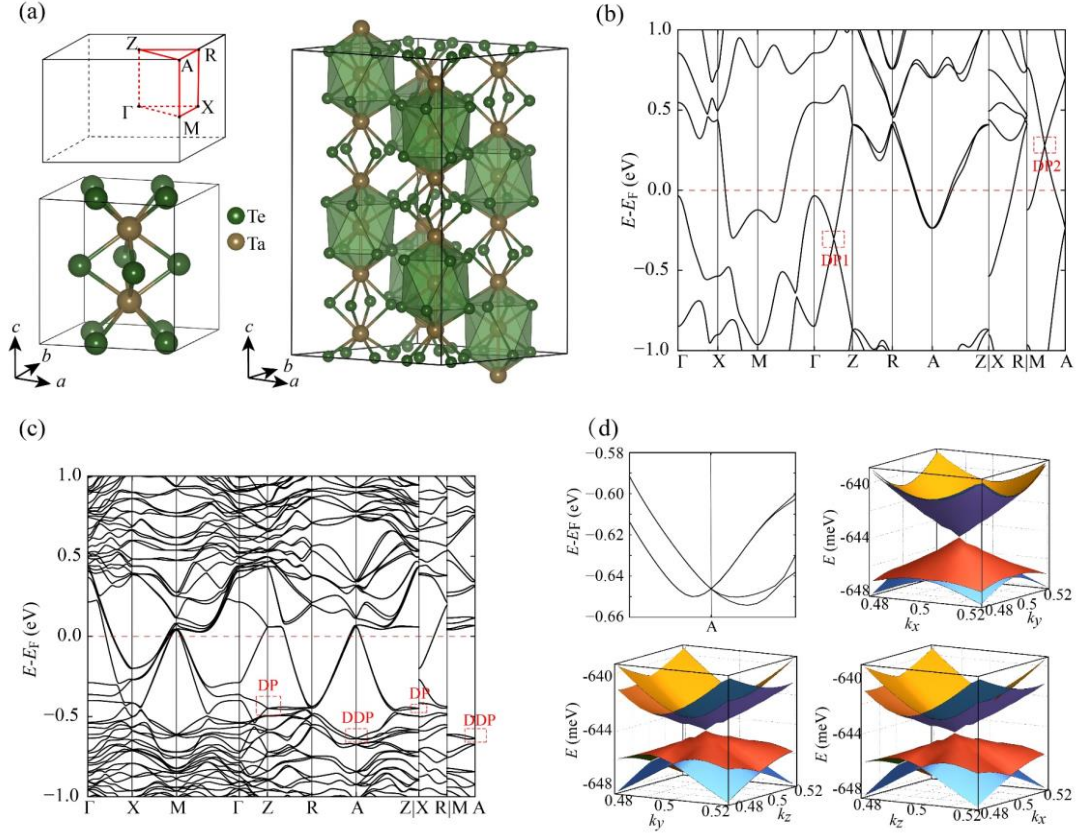

FIG. 1. Band structures of TaTe$_4$ with and without CDW. (a) Lower left panel and right panel shows unit cell of non-CDW phase and CDW phase TaTe$_4$, where green polyhedrons indicate Ta$_3$ clusters. Upper left panel is the 3D bulk Brillouin zone of non-CDW/CDW phase TaTe$_4$ with high-symmetry points indicated. (b) Band structure along high-symmetry lines in the non-CDW Brillouin zone of TaTe$_4$ without CDW. Red dashed squares indicate DPs. (c) Band structure along high-symmetry lines in the CDW Brillouin zone of CDW phase TaTe$_4$. Red dashed squares indicate DP and DDP. (d) Magnified dispersion around DDP in c and 3D linear dispersion around DDP. $x$, $y$ and $z$ respectively correspond to the direction of $a$, $b$ and $c$ axes.

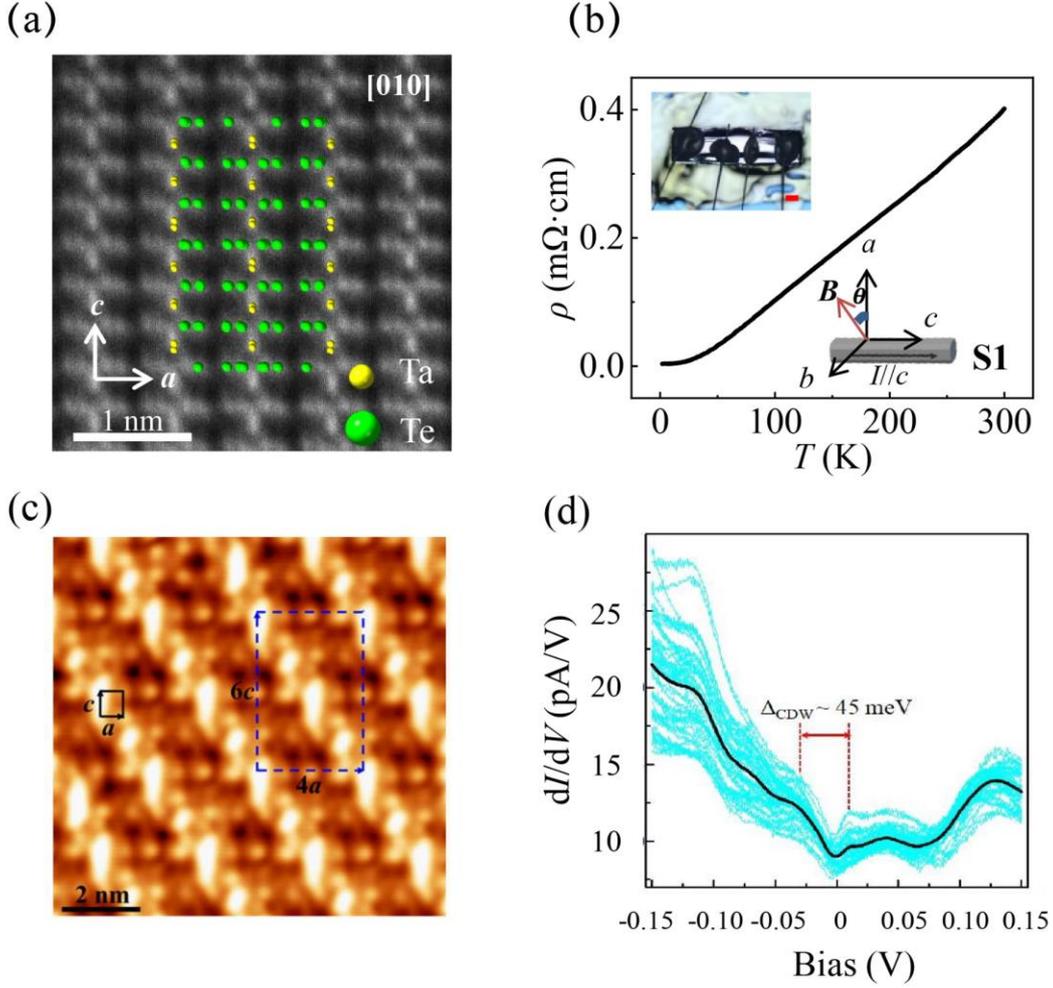

FIG. 2. Crystal structure, CDW and $\rho$ - $T$ property of TaTe$_4$. (a) HAADF-STEM image of TaTe$_4$ ac plane at room temperature with schematic structure. Scale bar represents 1 nm. (b) Sample resistivity as a function of temperature. Upper inset shows an optical image of one of our measured samples. Scale bar represents 200 μm. Lower inset is a schematic plot of magnetotransport measurement setup. (c) STM image of a cleaved ac surface with CDW modulation ($V_b$ = -250 mV, $I_t$ = 250 pA, image size is 10×10 nm$^2$). Blue and black squares represent surface unit cell with and without CDW, respectively. (d) STS in a region with CDW modulation. Cyan curves are numerous d$I$/d$V$ spectra measured at different locations on the bare surface areas (away from the defects) and the black curve is the averaged spectrum. A CDW gap of around 45 meV is clearly identified around Fermi level. Set point: $V_b$ = 150 mV, $I_t$ = 200 pA, the bias modulation for the lock-in technique is 9 mV.

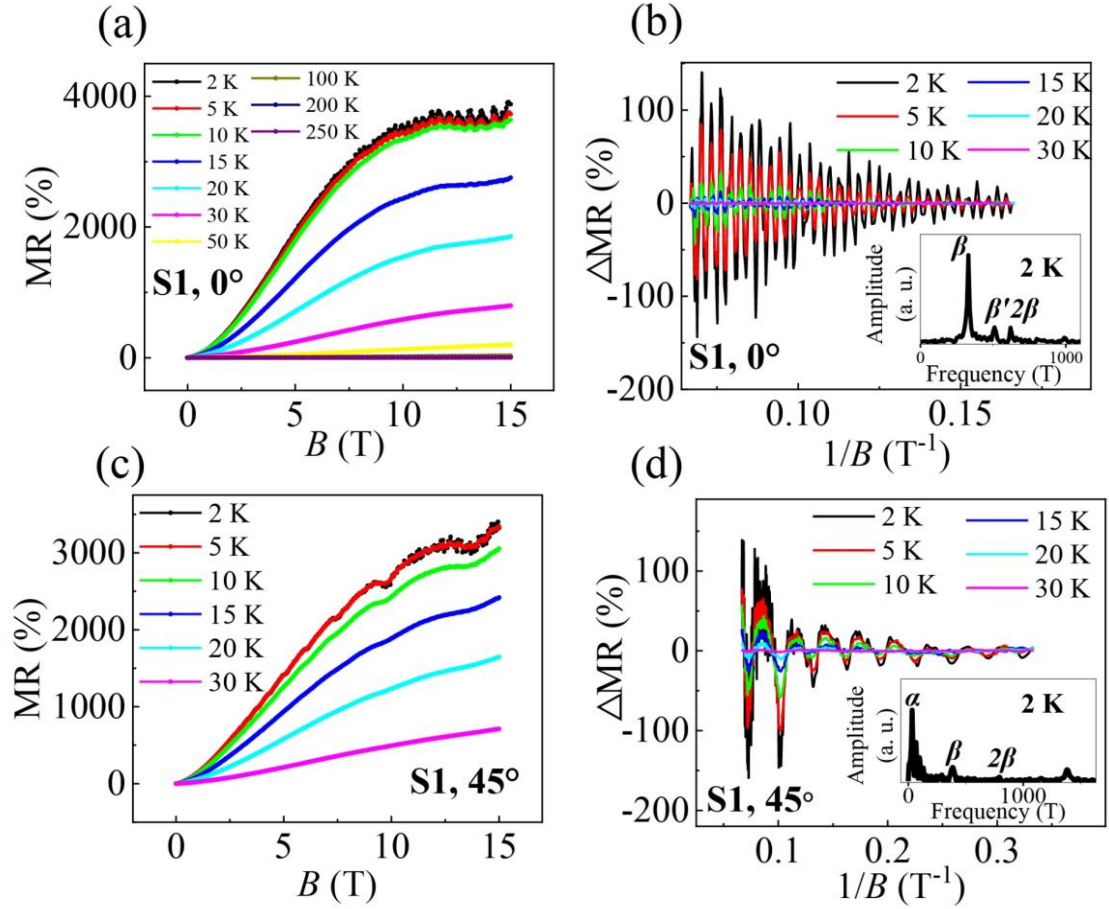

FIG. 3. Magnetoresistance and SdH oscillations in TaTe$_4$. (a)-(b), MR and SdH oscillation in TaTe$_4$ measured at $\theta = 0°$. (a) MR as a function of magnetic field applied along $a$ axis at various temperatures. (b) Oscillation component in (a) after subtracting a smooth background. Inset shows the FFT of oscillation at 2 K. (c) and (d) are results for $\theta = 45°$.

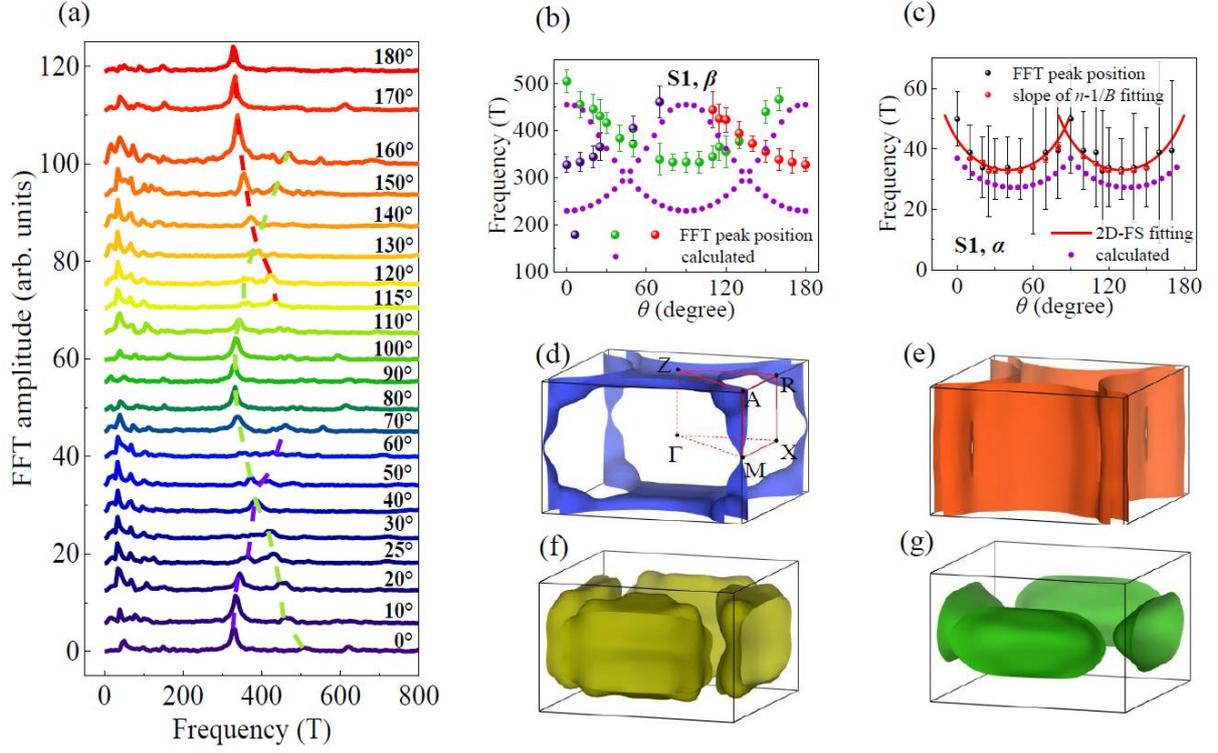

FIG. 4. Angular-dependent SdH oscillation and CDW phase Fermi surface of $TaTe_4$. (a) FFT of oscillation component obtained by subtracting an eighth-order polynomial background at various $\theta$. Dashed lines are guides to the eye. (b) Extracted high frequency peak positions in (a) as a function of $\theta$. Colors of data indicate contributions from different but equivalent pockets. Violet points are numerical results from calculated FS in (g). (c) Extracted low frequency peak positions in a and frequencies extracted from Landau level fan diagrams as a function of $\theta$. Violet points are numerical result from calculated FS in (d) and red lines are 2D-FS fitting to data. (d)-(g), The FS of band 1-4 which crosses Fermi level of CDW phase $TaTe_4$ in the first Brillouin zone.

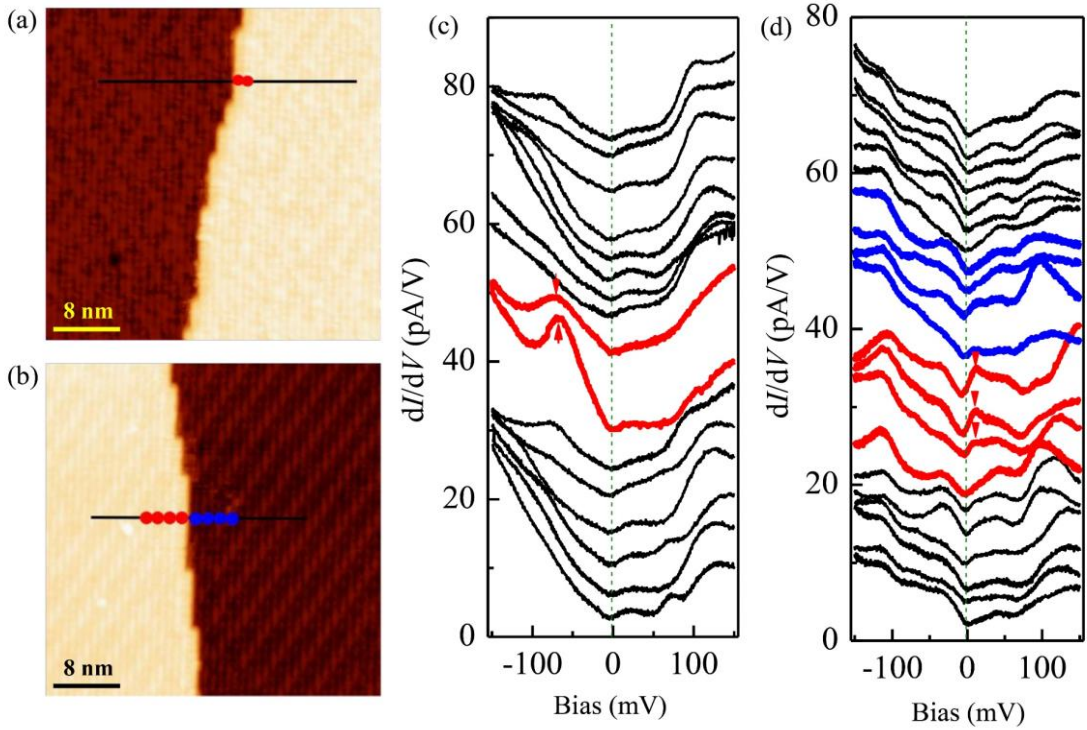

FIG. 5. States residing on step edges observed by STM. (a) STM image of an edge existing in a region with weak CDW modulation ($V_b$ = -70 mV, $I_t$ = 200 pA, image size is 40×40 nm$^2$). (b) STM image of an edge existing in a region with clear CDW modulation ($V_b$ = 12 mV, $I_t$ = 200 pA, image size is 40×40 nm$^2$). (c) Tunneling spectra at different locations along the line in (a). Spectra near the edge are shown in red. (d) Tunneling spectra at different locations along the line in (b). Spectra near the edge are shown in thick curves. Red arrows mark the appearance of edge states. All spectra were taken with the set point: $V_b$ = 150 mV, $I_t$ = 200 pA. The magnitude of the bias modulation for the lock-in technique is 9 mV.

**Supplementary Material for**

# Eightfold Fermionic Excitation in a Charge Density Wave Compound

Xi Zhang,[*] Qiangqiang Gu,[*] Haigen Sun,[*] Tianchuang Luo,[*] Yanzhao Liu, Yueyuan Chen, Zhibin Shao, Zongyuan Zhang, Shaojian Li, Yuanwei Sun, Yuehui Li, Xiaokang Li, Shangjie Xue, Jun Ge, Ying Xing, R. Comin, Zengwei Zhu, Peng Gao, Binghai Yan, Ji Feng,[†] Minghu Pan,[‡] Jian Wang[§]

[*]These authors contributed equally to this work.
[§]Corresponding author.
jianwangphysics@pku.edu.cn (J.W.)
[‡]Corresponding author.
minghupan@hust.edu.cn (M.P.)
[†]Corresponding author.
jfeng11@pku.edu.cn (J.F.)

**This file includes:**

    **Supplementary section**

I. Crystal Structure of $TaTe_4$ and Further Characterizations
II. STS of CDW Defect Regions
III. MR and SdH Oscillations Measured under 56 T Pulsed Magnetic Field
IV. Effective Mass and Dingle Fitting of SdH Oscillations up to 15 T
V. Hall Measurement of $TaTe_4$
VI. Angular-dependent SdH Oscillations of S1
VII. Fermi Surface and Angular-dependent SdH oscillation frequencies in non-CDW Phase
VIII. Landau Level Fan Diagram for $β$ Pocket

    **Figures S1-S11**

# I. CRYSTAL STRUCTURE OF TaTe$_4$ AND FURTHER CHARACTERIZATIONS

Figure S1 shows the crystal structure of the non-CDW phase of TaTe$_4$. Fig. S2 shows the simulated crystal structure of the CDW phase of TaTe$_4$, with the method presented in Methods section of the article.

Figure S3a shows the FFT of STEM image of TaTe$_4$ $ac$ surface. Spots representing enlarged unit cell under CDW is clearly observed. Fig. S3b shows XRD pattern of ($l$00) surface of TaTe$_4$ single crystal. Surfaces are labeled by CDW unit cell. As in Fig. 2c, bright strips observed under STM in Fig. S3c indicates CDW modulation. By performing FFT to Fig. S3c, spots corresponding to undistorted unit cell and enlarged unit cell under CDW can be observed [Fig. S3d]. Enlarged surface cell is thus determined to be $4a \times 6c$.

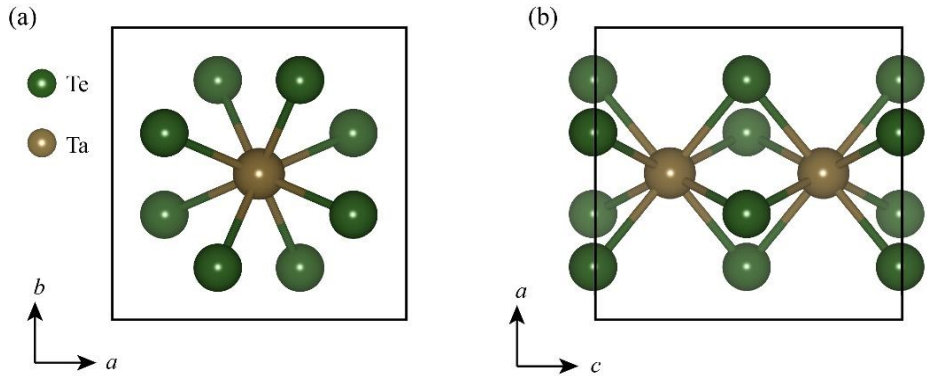

FIG. S1. Crystal structure of TaTe$_4$ in the absence of CDW. (a) Crystal structure of TaTe$_4$ in the absence of CDW when viewed from $c$ axis. (b) Crystal structure viewed from $b$ axis. (a) and (b) are visualized using VESTA [1]. The black rectangles show the unit cell. The Green and golden spheres represent Te and Ta atoms, respectively.

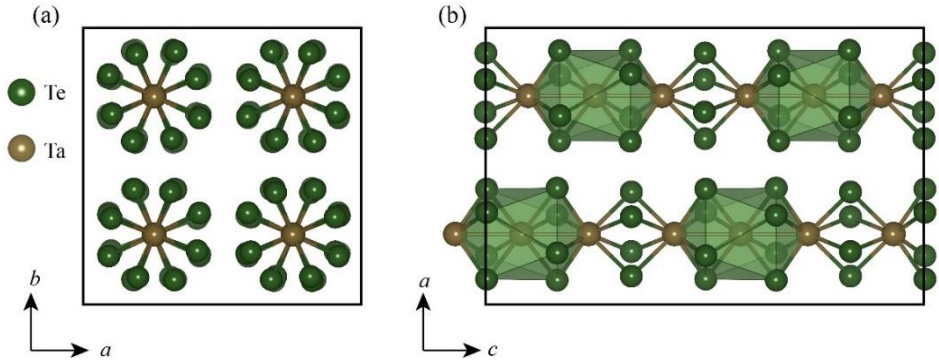

FIG. S2. Crystal structure of the CDW phase of TaTe$_4$. (a) Crystal structure of TaTe$_4$ CDW phase when viewed from $c$ axis. (b) Crystal structure viewed from $b$ axis. Green polyhedrons indicate Ta$_3$ clusters. (a) and (b) are visualized using VESTA [1]. The black rectangles indicate the enlarged $2a \times 2b \times 3c$ unit cell. The Green and golden spheres represent Te and Ta atoms, respectively.

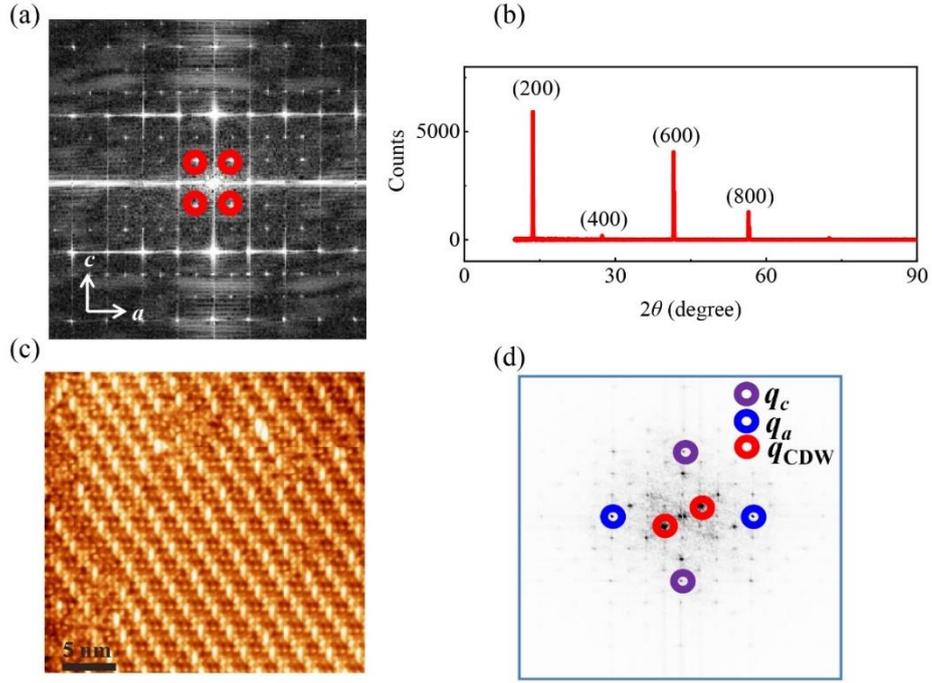

FIG. S3. Further characterization of TaTe$_4$ CDW phase. (a) FFT of the STEM image of TaTe$_4$ $ac$ plane. Red circles indicate FFT spots representing enlarged unit cell under CDW. (b) XRD pattern of TaTe$_4$ ($l$00) surfaces. The surfaces are labeled by CDW unit cell. (c) CDW modulated surface observed by STM ($V_b$ = -150 mV, $I_t$ = 250 pA, image size is 40×40 nm$^2$). (d) FFT of (c). Red circles indicate spots for CDW while blue and violet circles are spots for undistorted lattice.

## II. STS OF CDW DEFECT REGIONS

We observe CDW defects under STM, where lattice is intact but a uniform CDW modulation is absent. An example is shown in Fig. S4a. The STS spectra obtained in the region do not feature a sharp CDW gap around Fermi level, as is shown in Fig. S4b, which is in sharp contrast to Fig. 2d. This confirms that the gap observed is indeed induced by a CDW instability.

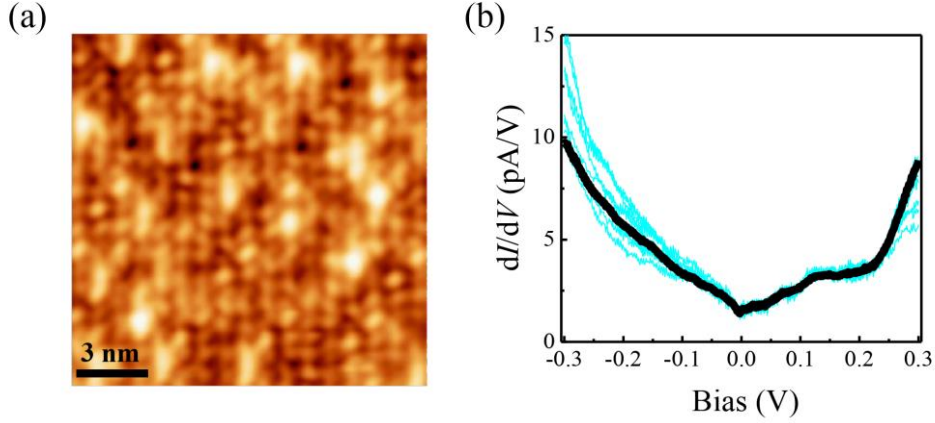

FIG. S4. STS of CDW defect. (a) Topographic image of a region without a uniform CDW modulation ($V_b$=-200 mV, $I_t$=100 pA, image size is 15×15 nm$^2$). (b) Numerous d$I$/d$V$ spectra measured at different locations on the regions without clear CDW modulation are shown in cyan and the averaged spectrum is shown in black. The CDW gap is reduced compared to Fig. 2d. Set point: $V_b$=150 mV, $I_t$=200 pA, the bias modulation for the lock-in technique is 2 mV.

## III. MR AND SdH OSCILLATIONS MEASURED UNDER 56 T MAGNETIC FIELD

We measure MR and quantum oscillation up to 56 T in a pulsed magnetic field for another sample (S2). The results are shown in Fig. S5a-c. At higher magnetic fields, two peculiar features are of interest. First, richer oscillation pattern is obtained. After performing FFT of the oscillation pattern, various peaks, including those with frequencies much higher than that observed up to 15 T, are clearly seen [Fig. S5d-f]. The peaks showing high frequencies correspond to SdH oscillations since their positions remain unchanged and the amplitudes decrease with the increase of temperature. Because they are superimposed on $\beta$ oscillation, further analysis of these SdH oscillations is hindered. These frequencies are likely to originate from magnetic breakdown between pockets, as is previously reported in other topological materials [2].

Another interesting feature is that magnetoresistance exhibits unconventional behavior with magnetic field above 15 T. For $\theta = 0°$, MR decreases after a threshold magnetic field and the threshold field increases with increasing temperature [Fig. S5a], indicating a low temperature negative magnetoresistance contribution. For $\theta = 90°$, MR increases linearly after a saturation below 15 T [Fig. S5c]. We analyzed $\beta$ oscillation measured by pulsed magnetic field. In order to exclude contributions from higher frequency oscillations, a band-pass filter is used to obtain clear, single-frequency oscillation [3], as is shown in Fig. S6a-c. Cyclotron mass obtained by fitting (Fig. S6d-f) gives $m_c =$ 0.35 $m_e$ (0°), 0.38 $m_e$ (45°), 0.37 $m_e$ (90°), which is consistent with that of S1 measured below 15 T.

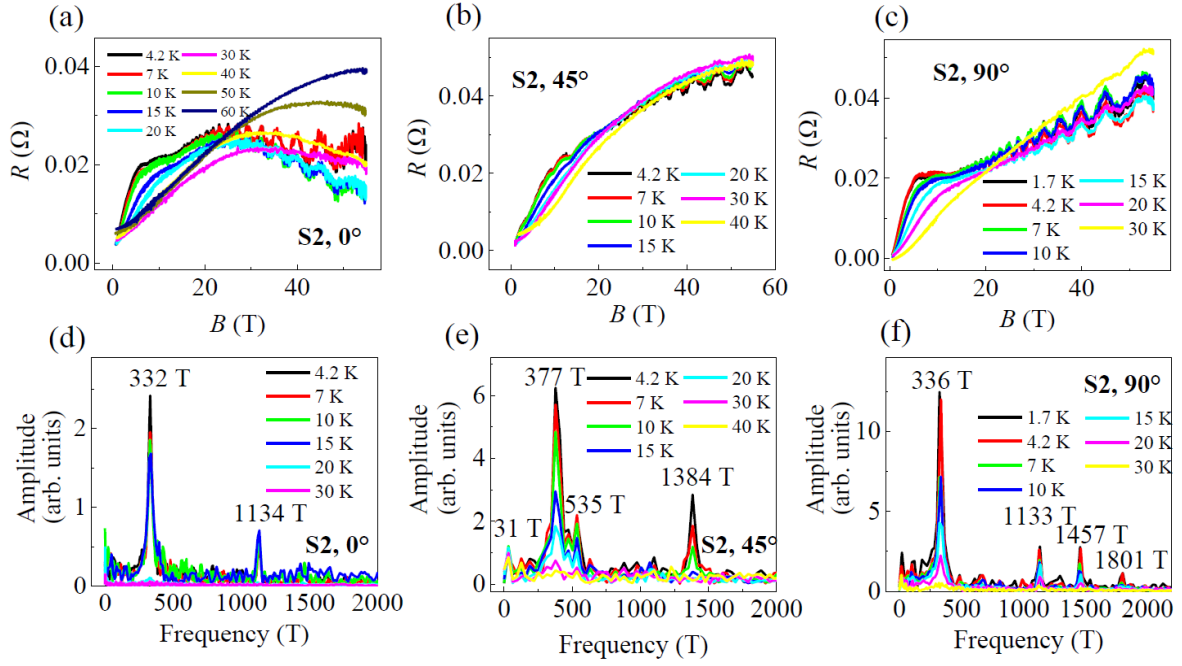

FIG. S5. MR and SdH oscillations measured by pulsed magnetic field up to 56 T. (a)-(c) Resistance as a function of magnetic field at various temperatures respectively for $\theta = 0°$, 45° and 90°. (d)-(f) FFT of corresponding SdH oscillation pattern. Labels indicate positions of FFT peaks.

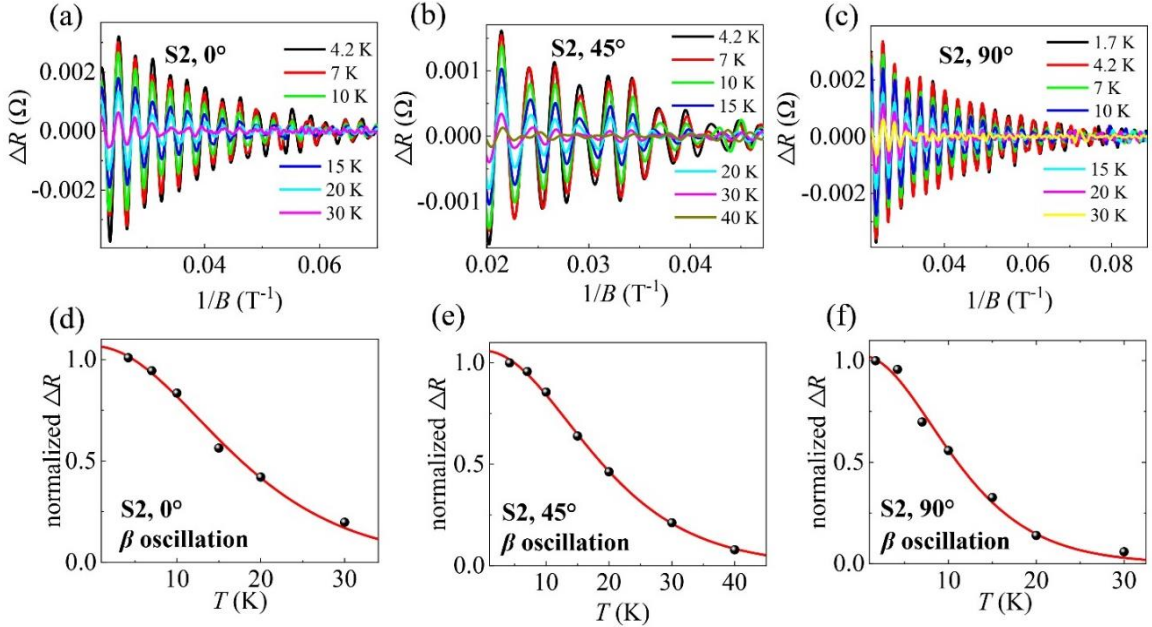

FIG. S6. $\beta$ oscillation in S2 measured by pulsed magnetic field up to 56 T. (a)-(c) $\beta$ oscillation in Fig. S5 obtained by band-pass filtering respectively for $\theta = 0°$, 45° and 90°. (d)-(f) Effective mass fitting corresponding to oscillation in a-c, which gives consistent result as that from $\beta$ oscillation measured below 15 T in S1.

## IV. EFFECTIVE MASS AND DINGLE FITTING OF SdH OSCILLATIONS UP TO 15 T

Fig. S7a and b show effective mass fitting as introduced in the main text. At a fixed magnetic field, oscillation amplitude $\Delta$MR damps with the increase of temperature and is quantitatively described by factor $R_L$ (main text). Effective mass can be obtained by direct non-linear fitting.

LK formula also includes damping effect due to scattering, which enables us to obtain quantum lifetime $\tau_Q$ from SdH oscillation pattern. From LK formula we have the following relation

$$\ln(\Delta \text{MR} \sinh(\chi) B) = -\frac{\pi m_c}{e \tau_Q} \frac{1}{B} + \text{const.} \tag{S1}$$

$\chi$ is defined in the main text. By performing such a linear fitting [shown in Fig. S7c and d], we can obtain $\tau_Q = 2.4 \times 10^{-13}$ s for $\beta$ oscillation at $\theta = 0°$ and $2.6 \times 10^{-13}$ s for $\alpha$ oscillation at $\theta = 45°$. It has been reported that this quantum lifetime is different from transport lifetime $\tau_{tr}$ in Drude model $\mu = \frac{e\tau_{tr}}{m^*}$ and the ratio $R_\tau = \frac{\tau_{tr}}{\tau_Q}$ should be larger than unity. For our sample, $\beta$ pocket has $R_\tau = 10$ for $\theta = 0°$. This result is comparable to that reported for GaAs-based two dimensional electron gas [4].

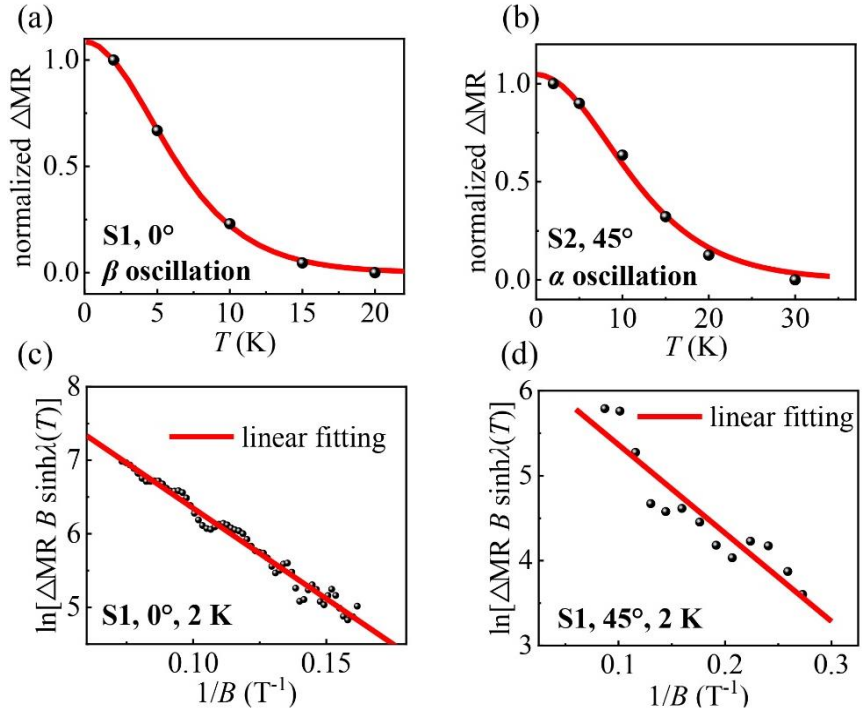

FIG. S7. Cyclotron mass and Dingle fitting of SdH oscillation in TaTe$_4$. (a)-(b) Cyclotron mass fitting of $\beta$ oscillation at $\theta = 0°$ and $\alpha$ oscillation at $\theta = 45°$. They respectively give $m_c = 0.33$ $m_e$ and $0.15$ $m_e$. (c)-(d) Dingle fitting of the two oscillations which yields quantum transport lifetime $\tau_Q$.

## V. HALL MEASUREMENT OF TaTe$_4$

Since TaTe$_4$ typically crystalize in a needle shape, hall measurement is rather difficult to perform. We measured longitudinal and Hall resistivity for sample S1 with standard 6-probe technique. The result of longitudinal resistivity is presented in the main text. After subtracting a symmetric component, we obtain anti-symmetric Hall signal measured at $\theta = 0°$. The data for positive magnetic field is shown in Fig. S8a. Up to 10 T, Hall resistivity is almost linear with magnetic field, indicating a dominant electron carrier. By linear fitting, we obtain carrier concentration as $n = \frac{1}{R_H e}$, where $R_H$ is the slope of linear fitting. The result is shown in Fig. S8b. As expected, carrier concentration increases with temperature. Low temperature carrier density is around $1.3 \times 10^{20}$ cm$^{-3}$. By Drude formula $\rho = \frac{1}{ne\mu}$, we can obtain carrier mobility at various temperatures [Fig. S8c]. At 2 K, carrier mobility is as high as $1.3 \times 10^4$ cm$^2$/(V·s). It reduces quickly with the increase of temperature. At 250 K, carrier mobility is about 20 cm$^2$/(V·s).

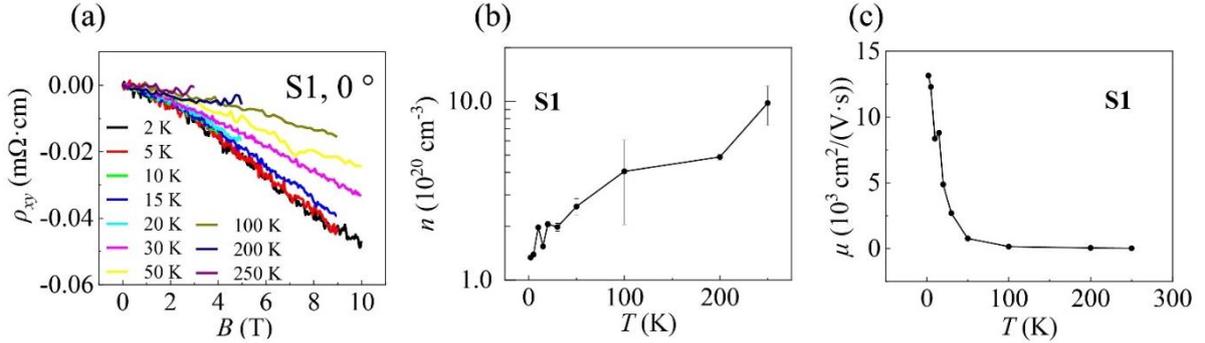

FIG. S8. Hall signal of TaTe$_4$. (a) Hall resistivity as a function of magnetic field along $\theta = 0°$ direction at various temperatures. (b) Carrier concentration obtained by linear fitting of data in a. (c) Mobility as a function of temperature obtained by carrier concentration and resistivity.

## VI. ANGULAR-DEPENDENT SdH OSCILLATIONS OF S1

Fig. S9 shows angular dependent SdH oscillation component of S1 obtained by subtracting eighth-order polynomial background. The FFT of Fig. S9 is presented in Fig. 4a.

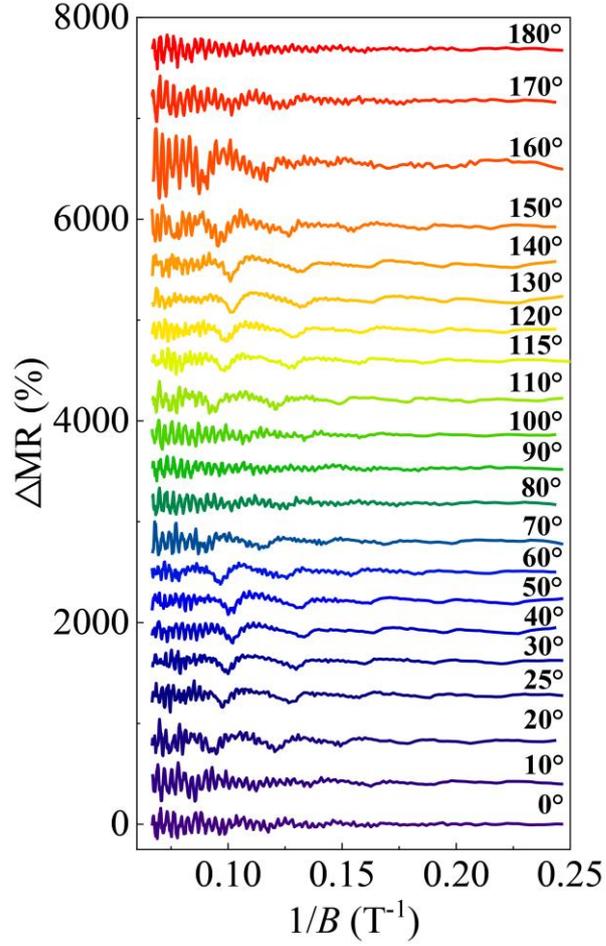

FIG. S9. Angular dependent SdH oscillation pattern at 2 K for S1 obtained by subtracting an eighth-order polynomial background.

## VII. FERMI SURFACE AND ANGULAR-DEPENDENT SdH OSCILLATION FREQUENCIES IN NON-CDW PHASE

We calculated the Fermi surface and possible SdH oscillation frequencies in the non-CDW phase [Fig. S10]. The experimentally observed frequencies cannot be explained by the non-CDW band structure.

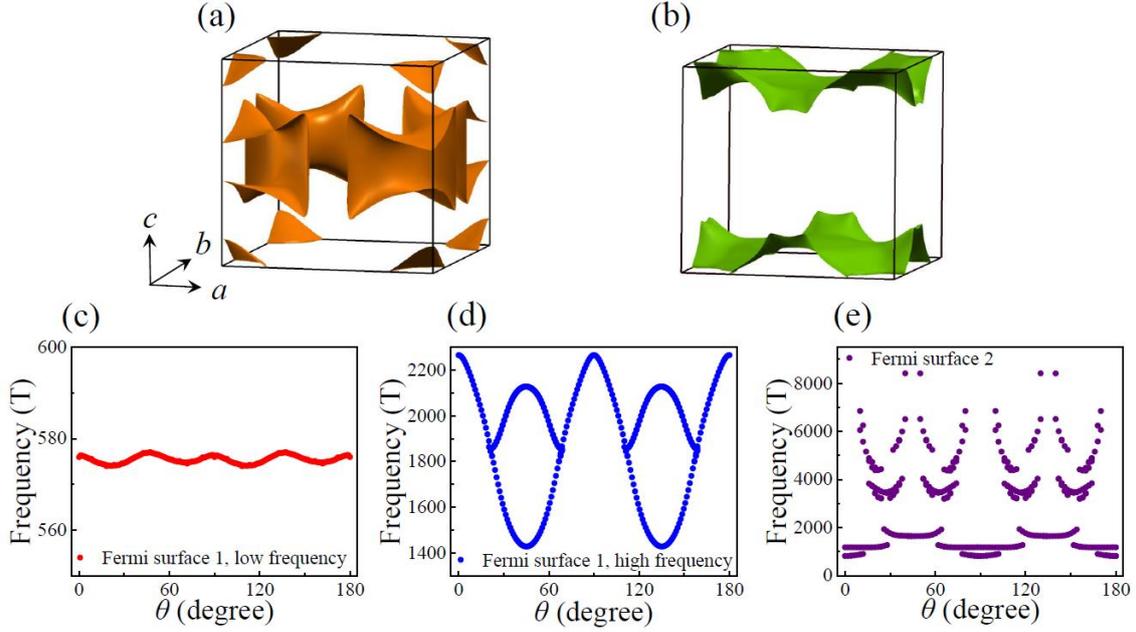

FIG. S10. Simulated Fermi surface and SdH oscillation frequencies in non-CDW phase of TaTe$_4$. (a)-(b) Fermi surface 1 and 2 in non-CDW Brillouin zone. (c)-(d) Possible SdH oscillation frequencies derived from Fermi surface 1. (e) Possible SdH oscillation frequencies obtained from Fermi surface 2. The frequencies are derived for **B** lying in the *ab* plane. $\theta$ is defined as the angle between **B** and *a* axis. The possible frequencies in the non-CDW phase cannot fit our experimental results.

## VIII.   LANDAU LEVEL FAN DIAGRAM FOR $\beta$ POCKET

Due to the large cross-sectional area of $\beta$ pocket, it is hard to reach low Landau index $n$ in our experimental condition. Although intercept obtained by Landau level fan diagram might then be not precise enough, we found out that several measured samples yield consistent value of intercept $\phi \approx -\frac{1}{8}$, traditionally regarded as corresponding to a nontrivial $\pi$ Berry phase. It is worth noticing that, since the double DP and the DP are far below the Fermi level in TaTe$_4$, the observed magnetic oscillations cannot provide related information. Previous studies on SdH oscillation frequently regard Landau Level (LL) fan diagram intercept $\phi$ as an indicator of Dirac or Weyl points [5-7]. However, by plotting LL fan diagram of $\beta$ oscillation at 0°, we consistently obtain a "nontrivial" value $\phi = -\frac{1}{8}$ in several samples [7] [Fig. S11]. This shows that factors besides band topology may also give rise to a "nontrivial" $\phi$ value. Thus, without further numerical or experimental evidences, $\phi$ value alone may not be justified to prove the existence of symmetry protected degeneracies.

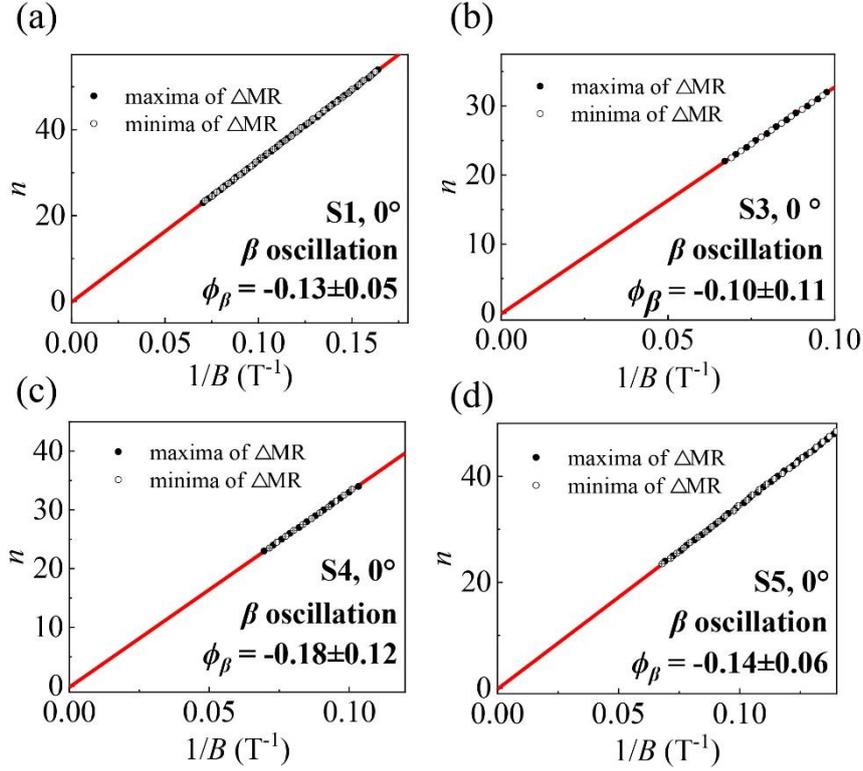

FIG. S11. Landau level fan diagram of $\beta$ oscillation in several samples at $\theta = 0°$. All of them give intercept value around $-1/8$.

---